\documentclass[ 
preprint, 
preprintnumbers,
 amsmath,amssymb,
 showkeys,
aps,
]{revtex4-2}

\usepackage{graphicx}
\usepackage{dcolumn}
\usepackage{bm}

\usepackage{url}
\usepackage{xcolor}

\newcommand{\scimed}{\textsc{SciMED}~}
\newcommand{\review}[1]{\textbf{#1}}

\makeatletter
\DeclareRobustCommand{\V}{\text{\volumedash}V}
\newcommand{\volumedash}{%
  \makebox[0pt][l]{%
    \ooalign{\hfil\hphantom{$\m@th V$}\hfil\cr\kern0.08em--\hfil\cr}%
  }%
}
\makeatother

\begin{document}

\preprint{}

\title{Improved Prediction of Settling Behaviour of Solid Particles through Machine Learning Analysis of Experimental Retention Time Data}

\author{Liron Simon Keren}
 \email{lirons.gb@gmail.com}
 \affiliation{Turbulence Structure Laboratory, Tel Aviv University, Israel}
\author{Teddy Lazebnik}
\affiliation{
 Department of Mathematics, Ariel University, Israel \\
 Department of Cancer Biology, Cancer Institute, University College London, UK
}

\author{Alex Liberzon}
\affiliation{
 {Turbulence Structure Laboratory, Tel Aviv University, Israel
}
}

\date{\today}

\begin{abstract}
The motion of particles through density-stratified interfaces is a common phenomenon in environmental and engineering applications. However, the mechanics of particle-stratification interactions in various combinations of particle and fluid properties are not well understood. This study presents a novel machine-learning (ML) approach to experimental data of inertial particles crossing a density-stratified interface. A simplified particle settling experiment was conducted to obtain a large number of particles and expand the parameter range. Using ML, the study explores new correlations that collapse the data gathered in this and in previous work by ~\citet{verso2019}. The "delay time", which is the time between the particle exiting the interfacial layer and reaching a steady-state velocity, is found to strongly depend on six dimensionless parameters formulated by ML feature selection. The data shows a correlation between the Reynolds and Froude numbers within the range of the experiments, and the best symbolic regression is based on the Froude number only. This experiment provides valuable insights into the behavior of inertial particles in stratified layers and highlights opportunities for future improvement in predicting their motion.

\end{abstract}

\maketitle

\section{Introduction}
\label{sec:intro}

Settling of inertial particles across layers of fluids of different densities appears in various engineering and environmental fluid mechanics problems~\cite{magnaudet2020, Mrokowska2018, Renggli2016MagmaSettling}. The settling velocity of particles can be estimated from the first principles only for cases of particles settling with low Reynolds ($Re_p = a V_p/\nu$, $a$ is particle \review{diameter}, $\nu$ fluid kinematic velocity) and low Stokes number \review{($St=T_p/T_f$, $T_p$ particle response time scale, $T_f$ flow response time scale)}. Furthermore, to be able to estimate the settling velocity, the particle must also settle through a fluid of homogeneous density ($\rho = \mathrm{const}$) or weakly linearly stratified fluids ($\partial \rho(z)/\partial z = \mathrm{const}$, where $z$ is fluid depth). When inertial particles cross interfacial layers, which are fluid layers with sharp density changes, there may be a noticeable amount of lighter fluid that follows the particle into interfacial layer with different densities. This lighter fluid is sometimes referred to as a "caudal wake." 
The coupled dynamics of the particle motion and of its caudal wake with respect to the surrounding fluid, together with the flow due to particle motion, lead to additional resistance on the particle. We will denote this additional resistance force by \(F_s\) as a single quantity, although there is an active discussion in the community about the origin, nature, and magnitude of various sources of the resistance force~\cite{magnaudet2020, verso2019, verso2020, Abaid2004, fernando1999, mandel2020}. Modeling accurately the time it takes for an inertial particle at a higher Reynolds and Stokes number to move through a stratified fluid with sharp density changes (hereinafter called interfacial layers) would be beneficial, for example, in problems of marine snow aggregations~\cite{MacIntyre1995, Prairie2013}, airborne or waterborne pollutant dispersion~\cite{Turco1983,kok2011, MacIntyre1995}, and oxygen levels regulation in the ocean~\cite{smith1992,burd2009}.  

Estimating the force components for different parameter regimes is extremely challenging due to the complexity of the dynamics of all the components, including the particle, the fluid layers, and the wake. Instead of integrating the force model in time, we suggest modeling the settling time directly, combining the experimental and machine learning methods. 

\subsection{Problem definition}\label{sec:definition}

\begin{figure}[ht]
    \centering
    \includegraphics[width=.75\textwidth]{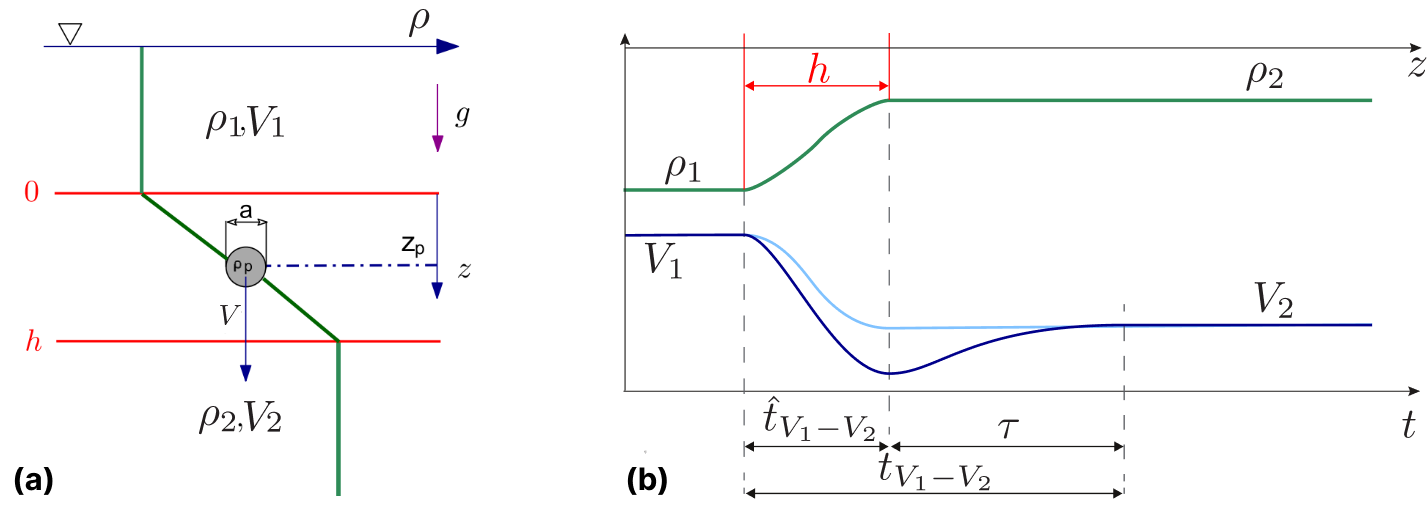}
    \caption{Schematic description of the problem of a particle crossing a stratified interfacial layer. The density of the fluid $\rho(z)$ changes vertically throughout the medium (green), from the first layer $\rho_1$ to the second layer $\rho_2$. The particle velocity $V(t)$ is illustrated for the case of negligible (cyan) or significant (blue) stratification force. The settling velocity in each layer is named \(V_1\) and \(V_2\), and the interfacial layer thickness is \(h\). We mark three relevant time intervals: the \review{theoretically predictable settling} time $\hat{t}_{V_1-V_2}$, the retention time $t_{V_1-V_2}$, and the delay time $\tau$.}
    \label{fig:problem_scheme}
\end{figure}

Let us consider the problem at hand schematically in Fig.~\ref{fig:problem_scheme}. In panel a), we show the scheme of the physical process: a particle heavier than the fluid is settling from the top, lighter fluid layer, through the interfacial layer of thickness $h$, and into the heavier fluid layer at the bottom (i.e., $\rho_1 < \rho_2 < \rho_p $). In panel b), we plot the fluid density profile of the two homogeneous fluid layers at densities $\rho_1$, $\rho_2$, and a continuous smooth transition $\rho(z)$ between the two homogeneous layers, called the stratified interfacial layer of thickness $h$ (green curve). We also plot the curve corresponding to the typical particle velocity expected to vary from $V_1$ to $V_2$ as a smooth monotonic function (cyan curve). This curve describes the case of a sphere settling without any additional resistance force stemming from the stratification, $F_s = 0$~\cite{magnaudet2020}. In some parameter regimes, the additional resistance is non-negligible, and there is possibly a non-monotonic change of velocity with a local minimum near the edge of the interfacial layer~\cite{fernando1999,verso2019,verso2020, mandel2020} (blue curve). 

In this study, we focus on the settling time estimate. Suppose the additional resistance force in the stratified interfacial layer is negligible. In that case, we will measure the theoretically predictable settling time of the particle, $\hat{t}_{V_1-V_2}$, defined as the time the particle moves from one \review{homogeneous-density layer} to another. Instead of settling time, we can determine the so-called retention time; the interval during which particle velocity changes from one terminal velocity value, $V_1$ to another terminal velocity value, $V_2$, marked as $t_{V_1-V_2}$. This definition is more beneficial for several cases, such as cases when the particle is heavier than both fluids $\rho_p > \rho_2$, when the particle is lighter $\rho_p < \rho_1$ and rises through the interfacial layer, as well as for the cases when particle temporarily changes its direction of motion and levitates ~\cite{Abaid2004, verso2019}.

For the monotonic, theoretically predictable case, the \review{theoretically predictable} settling time through the interfacial layer is equivalent to the retention time \review{(i.e., $\hat{t}_{V_1-V_2}=t_{V_1-V_2}$}. Suppose there is an unknown additional resistance force. In that case, the retention time is longer because it also contains the interval during which particle settling velocity is lower than both the steady-state values or it levitates and changes the velocity sign. We define the difference between the expected \review{settling} time and observed retention time as the ``delay time'', marked as $
\tau$ (see Fig.~\ref{fig:problem_scheme}). In this study, we develop the method to predict $\tau$ based on particle and fluid properties, using a machine learning model trained on experimental data. 

\subsection{Existing data} \label{sec:literature_survey}

\begin{figure}[ht]
    \centering
    \includegraphics[width=1\textwidth]{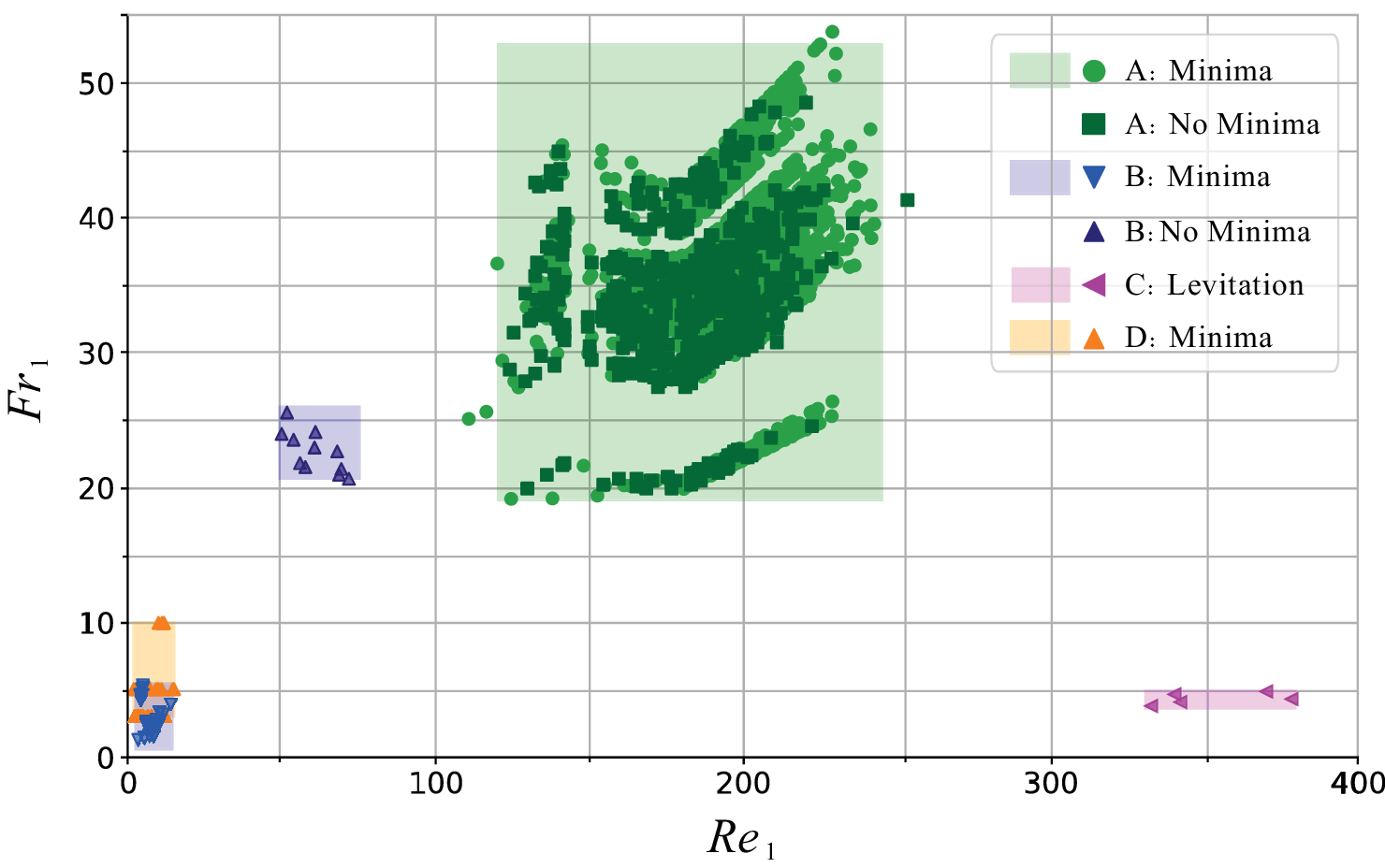}
    \caption{Summary of the parameter range of the experimental results of \citet{verso2019, Abaid2004, fernando1999} and this study, in terms of entrance Froude versus entrance Reynolds numbers. A - present study, B - \citet{verso2019}, C - \cite{Abaid2004}, and D - \cite{fernando1999}. \review{Minima - refers to particles that exhibited a clear local minimum velocity, No minima - refers to particles without a local minimum, and Levitation - refers to particles that momentarily reverse their motion upwards, against gravity.}}
    \label{fig:Re_Fr_map}
\end{figure}

Only a few studies address the problem of inertial particles crossing sharp interfacial stratified layers ($h/a \sim O(10)$, \review{$h$ is the interfacial layer thickness, and $a$ is the particle diameter}) between two miscible fluids of densities $\rho_1$ and $\rho_2$. The key parameters are the ``entrance'' Reynolds number, defined with the particle size and the velocity and viscosity of the layer from which the particle enters the interfacial layer: $Re_1 = V_1 a/\nu > 10$, and the corresponding entrance Froude number,  $Fr_1 = V_1 /N a < 100$, where $N$ is calculated as defined below in Eq.~\eqref{eq:Brunt-Vaisala}. In Fig.~\ref{fig:Re_Fr_map}, we summarize the existing results on the map of $Re_1, Fr_1$. \review{The figure also demonstrates the limited number of studies in the literature that referred to this parameter range, in which $Re_1 > 10$, i.e., the particles have significant inertia, and at the same time, the stratification is relatively strong, i.e., $Fr_1 < 100$. Outside of this parameter range, there are many more studies and comprehensive results. Most of these results are for cases of linear stratification and small particles at creeping flow regime, where $Re_1 < 1$}

\review{In the suggested parameter regime,} the first experimental study of inertial particles settling through an interfacial layer of finite thickness is by~\citet{fernando1999}. The authors used a water-alcohol-brine system and particles in the ranges $1.5 < Re_1 < 15$, $3 < Fr_1 < 10$. The authors attributed the particle slowdown to the additional drag force due to caudal fluid. In addition, they mentioned the plausible contribution of internal waves or modification of flow structure around the particle due to the density gradient~\cite{fernando1999}. However, the authors studied only the time it took the spheres to reach minimum velocity and did not investigate the phenomena in the denser bottom layer after the particles crossed the interface. Therefore, we do not have their estimate of the \review{retention time (as it requires the tracking of the particle within the bottom layer), and thus cannot infer} the particles' delay time $\tau$. \review{This prevents us from using their measurements to train or validate our model}. 

\citet{Abaid2004} performed similar experiments with sharper interfaces in the ranges of $1 \leq h/a \leq 5$, $ 20 < Re_1 < 450$, and $ 5 < Fr_1 < 20$. Note that the range does not mean that authors changed parameters systematically; it only marks the minimal and maximal values. In Fig.~\ref{fig:Re_Fr_map}, the pink rectangle marks the parameter range of the reported measurements of~\cite{Abaid2004}. Despite the similar range of Reynolds numbers to those of \cite{fernando1999}, the authors observed different caudal wake effects and reported a ``temporal levitation'' phenomenon, where spheres momentarily reverse their motion upwards, against gravity. \citet{Abaid2004} formulated a closure model as an additional virtual mass with its own degree of freedom, mimicking a caudal wake that moves upwards when the particle moves downwards. The authors did report prolonged time periods until the sphere regained steady state motion in the bottom layer, but did not study any time scales. 

\citet{verso2019} estimated the retention time based on the results of an experiment similar to \citet{fernando1999}. The authors used a water-alcohol-brine system in ranges of $2<Re_1<14$, $0.6<Fr_1<4$, and $h/a \sim 10$. \citet{verso2019} did not observe the levitation phenomenon in their parameter range. However, they proposed a model for the additional resistance due to stratification and caudal fluid that helped to estimate the parameter range of $Re, Fr$, and $h/a$ in which the levitation is possible. In addition,~\citet{verso2019} observed very prolonged $t_{V_1-V_2}$ timescales and developed their parametric model for the stratification force $F_s$. The authors also show that the crossing time $t_{V_1-V_2}$ is inversely proportional to the particle Reynolds number in the bottom layer, $Re_2= V_{2} a/\nu_2$. Their parametric model also predicted the data from the literature~\cite{fernando1999}. This is the only data that we could use for training, along with the data measured in our experiments. 

Recently, \citet{Wang2023} reported in their preprint a detailed experiment on the bouncing effect of particles, similar to the levitation reported by \citet{Abaid2004}. The authors measured and numerically simulated velocity fields and suggested a model of the resistance force. Unfortunately, their data is not yet available for comparison. 

The studies mentioned above are based on detailed measurements of a relatively low number of particles, $5 - 50$, mainly due to the technically challenging experiments. The major challenges relate to control of the thickness and location of the interface without mixing the fluid layers, handling small spheres ($a \sim 10 \div 10^3 \; \mu$m), tracking the particles with high spatial resolution, and in some cases enforcing refractive index matching. Although these experiments provide insight into the fluid mechanics and dynamics of the particles inside the interfacial layers, they do not create a statistically sufficient dataset to model the delay times. 

We also suspect that there are more useful forms of Reynolds and Froude numbers using different velocity and length scale combinations. It appears that the present typical distinction of parameter regimes used in the literature in terms of $Re_1$ and $Fr_1$~\cite{magnaudet2020,blanchette2012,Doostmohammadi2014,Yick2009,Camassa2010} is incomplete. Examination of the map presented in Fig.~\ref{fig:Re_Fr_map} raises questions about its specificity regarding the effect of stratification resistance. We could expect a map on which it is more clear which particles experience different physics (feeling a significant caudal wake resistance) and which do not (i.e., crossing with quasi-steady-state velocity values). 

In this study, we propose another approach: we deliberately simplify the experimental design, avoiding the difficulties associated with estimating the force component, $F_s$, \review{particularly the need} for index refraction matching. \review{Note that the correlation between Fs and index refraction stems from the need to precisely track the particles within the interfacial layer to measure Fs. This entails meticulously matching the refractive indexes of the top and bottom layers, ensuring minimal refractive differences throughout the particle's trajectory. By avoiding the need to match the refraction index of the top and bottom fluid layers}, we can \review{simplify the experimental design and} significantly extend the parameter space to include previously unexplored regimes. 

Furthermore, the simplified experimental design allowed us to obtain an unprecedented number of particle trajectories. We use this sufficiently big data set in the unexplored parameter regime with the custom-designed ML-based symbolic regression tool, \scimed~\cite{scimed}. We developed this tool to find symbolic regression correlations of $\tau$, using hidden non-dimensional parameters, that might have the potential to better explain the underlying physical mechanisms. This approach leads to an opportunity to find a new parametric predictive model for $\tau$, using a ``data-driven'' methodology~\cite{scimed}. 

\begin{figure}[ht]
    \centering
    \includegraphics[width=1\textwidth]{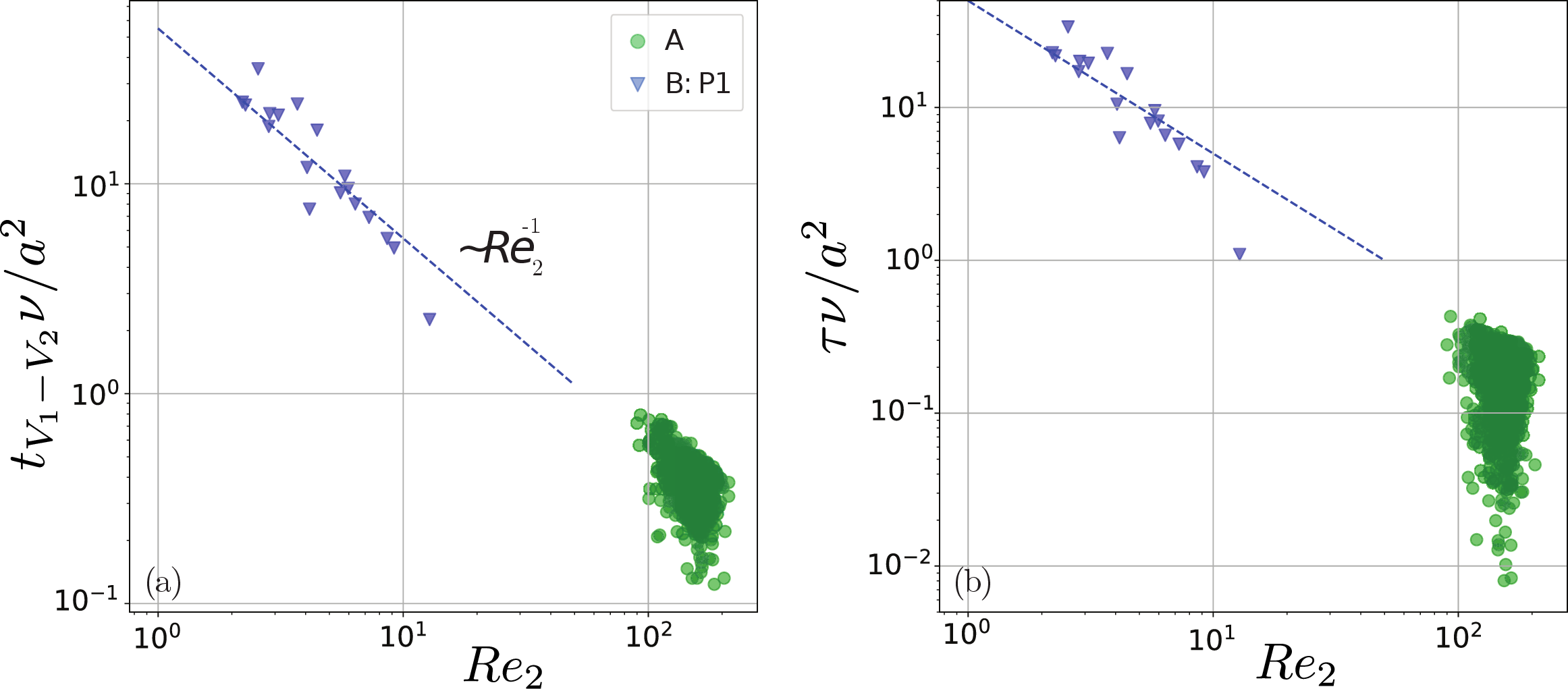}
    \caption{Normalized recovery time (a) and normalized delay time (b) versus $Re_2$. The graphs include measurements collected during our experiments (A) and those of particle type 1 from \citet{verso2019} (B).}
    \label{fig:Versos_corr}
\end{figure}

In Fig.~\ref{fig:Versos_corr}, we plot our data together with the only data and existing correlation for the delay and retention times by~\citet{verso2019}. It is clear that we arrived at a different parameter regime for which previous correlations do not match.

\section*{Materials and Methods}\label{sec:materials}
\begin{figure}
    \centering
    \includegraphics[width=0.8\textwidth]{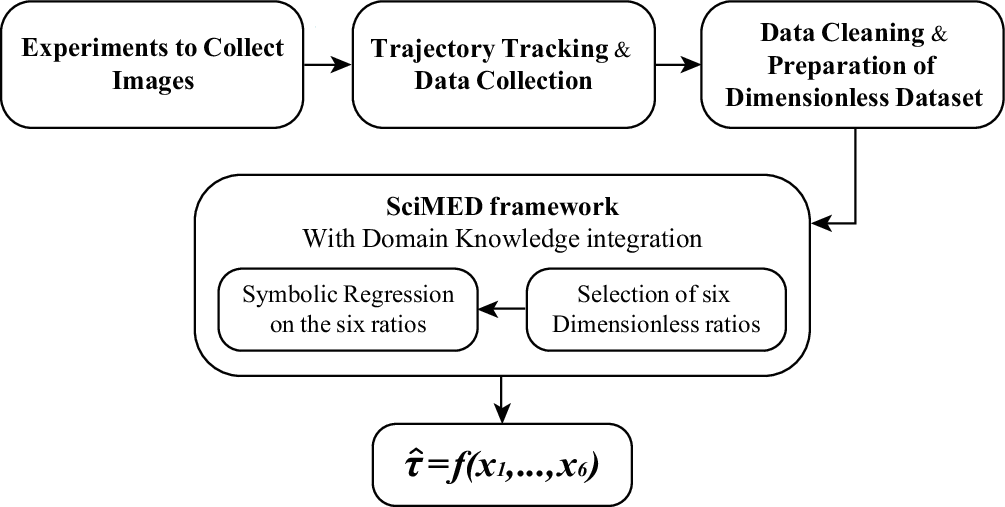}
    \caption{Scheme of the methodology applied in this research.}
    \label{fig:method_scheme}
\end{figure}

In this section, we describe in depth the materials, equipment, measures, and analysis techniques that were used, following the order of the scheme presented in Fig.\ref{fig:method_scheme}. The unprecedented number of particle trajectories collected enables the use of machine learning methods. For that reason, we developed and applied a specific method that we abbreviated \scimed, described in detail in the recent publication~\cite{scimed}. Our main interest at the end of the process is to find a new correlation that fits the available data of $\tau$ as a function of particle and fluid parameters.

\subsection*{Experimental Setup}
We conducted experiments in a glass tank with a cross-section of \(200\times 200 \) mm$^2$ and a depth of \(300\) mm, creating a top lighter layer of water (marked as layer 1) above a bottom heavier layer of water and Epsom salt (MgSO4) solution (marked as layer 2). We first fill the lighter fluid and pump the heavy fluid from a valve at the bottom of the tank, resembling the method used in~\citet{verso2017,verso2019}. This results in an interfacial stratified layer between two fluid layers, created through molecular diffusion. The interfacial layer is growing very slowly with time (less than 10 mm per day), and in the present experiments, it is in the range \(h = 10a \div 100a \). 
We varied the ratio of densities of the fluid layers $\rho_2/\rho_1$ in the range of 1.040 to 1.200. For these low concentrations of Epsom salts, it is customary to ignore the small variation in fluid viscosity. Alike previous studies~\cite{fernando1999}, we assume equal kinematic viscosity of both layers, $\nu_1=\nu_2 = \nu = 10^{-6}$ m$^2$ s$^{-1}$.  
The values of fluid density (\(\rho_1, \rho_2\)) and Brunt-Vaisala frequency (\(N\)) are shown in Table~\ref{tbl:fluid_properties}, where in this study, $N$ is calculated as follows: 
\begin{equation}
\label{eq:Brunt-Vaisala}
N=\sqrt{\frac{2 g}{\rho_1+\rho_2} \frac{\rho_2-\rho_1}{h}}.
\end{equation}
\begin{table}[!ht]
\centering
\caption{Ranges of the properties the fluid layers varied in, across all experiments.}
\label{tbl:fluid_properties}
\begin{tabular}{ p{2.5cm}p{2.5cm}p{2cm}p{2cm}p{2cm}p{2cm}p{cm} }
\hline
$\rho_1$ & $\rho_2$ & T & $\hat{h}$ & $N$ & $\nu$ \\
(g cm$^{-3}$) & (g cm$^{-3}$) & ($^\circ C$) & (cm) & (s$^{-1}$)& (m$^2$ s$^{-1}$) \\
\hline
0.997 - 1.002 &  1.038 - 1.200  &   16 - 23.5 & 2.94 - 5.11  &  3.05 - 6.51 &  $1\times 10^{-6}$ \\
\end{tabular}
\end{table}

Both the upper and lower fluid layers of the resulting medium are sufficiently deep for particles to reach terminal velocity. We used one type of high-quality commercially available spherical particles (Cospheric Inc., Santa Barbara, CA). These \review{smooth} spherical polystyrene particles have a density of 2.5 g cm\(^{-3}\) and a diameter range of 1000–1180 \(\mu\)m. We verified particle parameters using microscopy imaging and custom image processing code. We have found that the particles are close to perfectly spherical, with a diameter that is normally distributed within the prescribed range.

We released particles at the center of the tank, below the free surface, using a syringe pump and a long glass pipette filled with water and particles. The diameter of the pipette exit is slightly larger than the particle diameter, ensuring that only single particles can exit. We controlled the rate of the syringe pump and ensured that there was a sufficient time interval between the releases. Thus the majority of the settling particles crossed an undisturbed steady-state interface. Sometimes, however, we were unable to prevent particles from concentrating at the exit of the pipette and exiting with insufficient time intervals between them. In these cases, particles settle one after another with vertical distances of about 10 - 100 diameters. The insufficient time between two consecutive particles means that some particles enter the interface at the same horizontal location, and the interface itself is distorted by the previous particle.

We repeated the process 17 times, with each experiment taking between 3 and 5.5 hours (from the moment of filling the tank until the moment of the last recorded measurement). We filmed the motion of the settling particles using a digital high-speed camera (Photron Nova, 1024 $\times$ 1024 pixels) with a frame rate of 250 frames per second.

\subsection*{Trajectory tracking and data collection}
\begin{figure}[ht]
    \centering
    \includegraphics[width=1\textwidth]{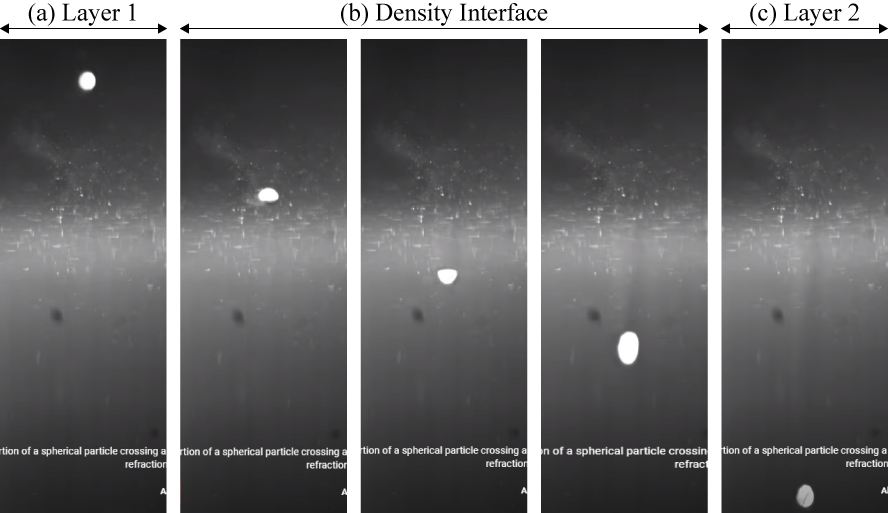}
    \caption{Digital snapshots of particle positions taken at different moments throughout its motion, moving from left to right sequentially. Spherical particle shapes in the left and right images indicate that the particle is currently settling through a homogeneous fluid, while the smeared elliptical shape in the middle images implies that the particle is in the virtual interface layer $\hat{h}$.}
    \label{fig:optical_distortion}
\end{figure}

We tracked the location of particles using the well-known particle tracking algorithm of OpenPTV~\cite{openptv, verso2019}. As our experiments were conducted with a relatively sharp interfacial layer, they have a significant drawback of substantial variation of refractive index in the interfacial layer. As a result, images of small particles from this region are significantly distorted, as demonstrated in zone (b) in Fig.~\ref{fig:optical_distortion}. Similar to other studies~\cite{alahyari1994particle}, it is impossible to obtain particle center in this region with an uncertainty smaller than a particle radius. Therefore, we could not measure particle velocity within the interfacial layer and focused only on measuring the information relevant to the delay time. Thus, we measured the time instant at which each particle starts slowing down, slightly before it enters the interfacial layer and the time instant at which it reaches the terminal velocity at the bottom layer, far below the layer of the strong refractive index gradient. 

\begin{figure}[ht]
    \centering
    \includegraphics[width=1.\textwidth]{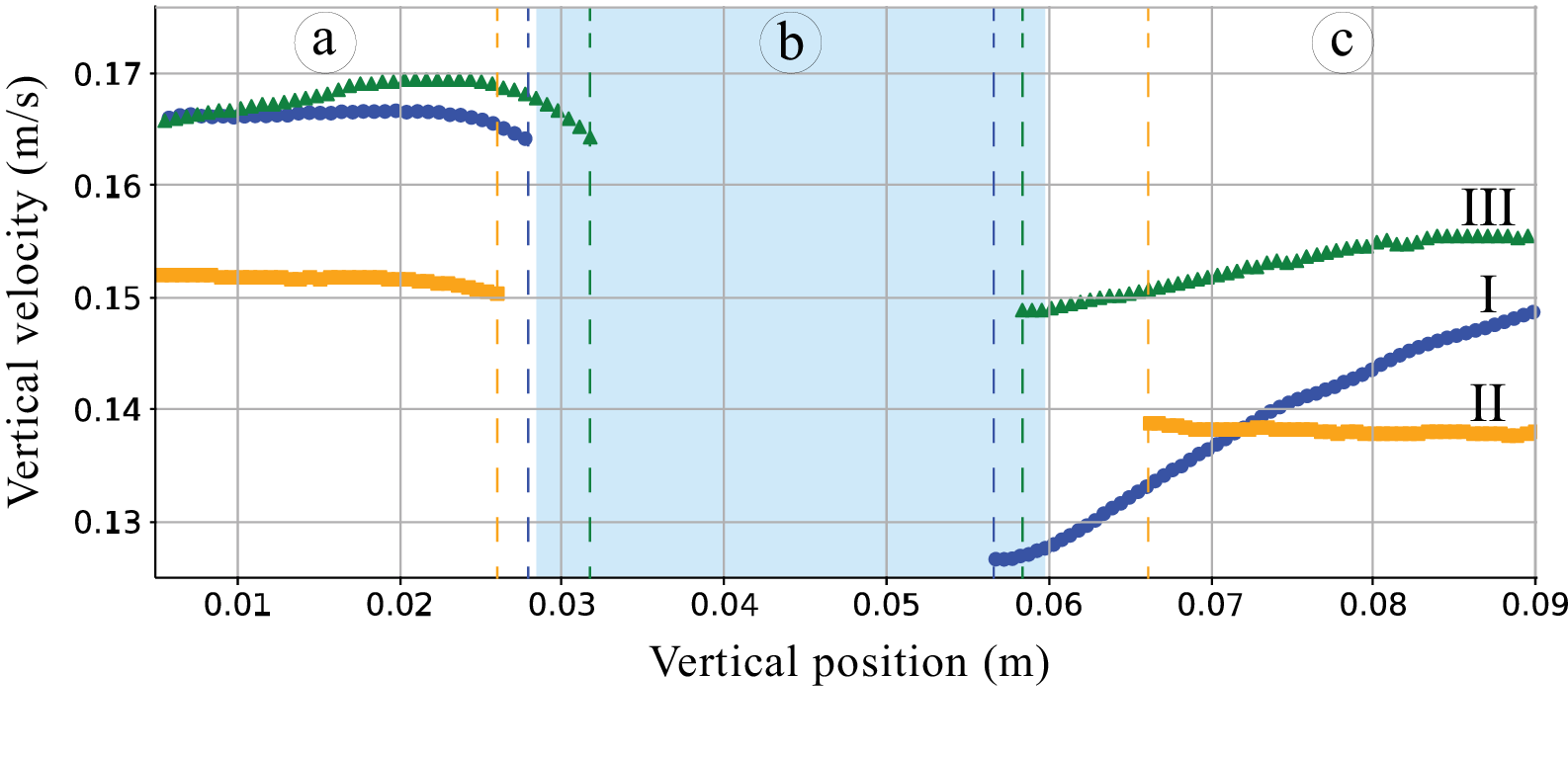}
    \caption{Three typical plots that can be found in the dataset, showing the dimensional velocity of a particle versus the distance traveled from the top of the viewing. These particular trajectories are taken from the same group of particles that were released successively in the presence of a density ratio of $\rho_2 / \rho_1 = 1.040$. The dashed lines indicate the stratification region as determined by each particle's motion, while the colored area (zone b) indicates the interfacial layer determined by the average of the located entrance and exit points. Zones marked as a-c indicate the upper and lower fluid layers.}
    \label{fig:typical_trajs}
\end{figure}

In addition, we use the particles to estimate the position of the interface. We mark the entrance and exit points for each trajectory based on particle velocity (outside the region of strong refractive index gradients). The entry point is marked at the location where $V(z)=0.99V_1$, and the exit point where either the particle reaches its minimum velocity, or if it doesn't experience a minimum, the terminal velocity $V_2$, similar to Refs.~\citep{Abaid2004,fernando1999,verso2017,verso2019}. We average these position values for groups of particles that were released in the same run and use this estimate as the approximate interface thickness $\hat{h}$. \review{In other words, $\hat{h}$ is an average of individual interface thickness measurements, collected from trajectories of particles that were released in proximity to one another.}
Fig. \ref{fig:typical_trajs} demonstrates the individual entrance and exit points (dashed lines) of three archetypes of particle trajectories that we measured, alongside the interface thickness $\hat{h}$ calculated based on them (blue rectangle). To the best of our knowledge, this is the first time this method for estimating the position and thickness of the interface has been used. Note that it is as accurate as other definitions of interface thickness, such as sampling of fluids at fixed depths, intrusive measurements with a conductance probe, or imaging methods using dye or Schlieren optical methods~\cite{fernando1999, Wang2023}. Our method avoids mixing and significantly prolongs the experimental run time, measuring thousands of trajectories. 

As seen in Fig.~\ref{fig:typical_trajs}, the three archetypes of particle trajectories are particles with a clear local minimum velocity ($I$), particles without a minimum ($II$), and particles that increase their velocity before entering the interfacial layer ($III$). 
Due to our inability to define particle position in zone (b), we ``tracked'' the particles in the sense that they are still linked to the same object, but there is no useful quantitative information in this region. Therefore, we masked the values tracked within the individually measured interface and used trajectories only to obtain $V_1$, $V_2$, the position of the minimum velocity, and the actual retention time $t_{V_1 - V_2}$. Specifically, to determine $V_1, V_2$, we averaged the first and last 20-100 measurement points (at 250 fps), with the number of measurements to be averaged depending on the position of the interfacial layer entrance in the frame. For trajectories of archetype $III$, we adjusted the number of averaged measurement points to avoid the time with increased velocities. For trajectories of archetype $I$, where the trajectory fails to stabilize within the limits of the frame, we determined $V_2$ as the average of the last five measurement values of each trajectory. Moreover, the retention time is determined by the time interval between the moment the particle returns to a terminal velocity by reaching $0.99 \leq V(t)/V_2 \leq 1.01$ and the time instant of departing from $V_1$ before entering  the interfacial layer. 

Next, similar to previous work~\cite{verso2019}, we estimate each particle diameter and density using the measurements of its terminal velocities in the top and bottom layers, and applying a force balance between the immersed weight, $F_{WB} =(\rho_p -\rho_f)\V g$, and the drag $F_D = C_D \frac{1}{2}\rho_f |V|VA$. Here $\rho_f$ is the density of the surrounding fluid (either $\rho_1$ or $\rho_2$ in the homogeneous layers), $\V=\pi a^3 / 6$ is the particle's volume, $A =\pi a^2 / 4$ is the projected surface area, and $C_D$ is the drag coefficient. The drag coefficient is estimated using the correlation~\citep{white1974viscous,verso2019}: 
\begin{equation}
C_D = 0.4 + \frac{24}{Re} +\frac{6}{1+\sqrt{Re}}
\end{equation}

We estimate the diameter and particle density using constrained optimization within the range provided by the manufacturer using the force balances in the upper and lower layers simultaneously, similar to Ref.~\cite{verso2019}.

In analogy to previous studies looking for correlations~\cite{fernando1999, Abaid2004,verso2019,boetti_verso_2022}, we also used two average properties: the average density of the fluid medium $\overline{\rho}$, and the average of terminal velocities $\overline{V} = 0.5(V_1 + V_2)$. From these properties, we are able to calculate the expected ``crossing time'' $\hat{t}_{V_1-V_2}=\hat{h}/\overline{V}$ that represents the time it takes the particle to cross $\hat{h}$ assuming that $F_s$ is negligible. Finally, we calculated the delay times $\tau$ as the delta between the actual retention time $t_{V_1 - V_2}$ and the expected crossing time $\hat{t}_{V_1-V_2}$ ($0\leq \tau < \infty$).

\subsection*{Data cleaning and preparation of non-dimensional dataset}
Footage of 3,264 settling particles is freely available as open access data, divided into 16 experiment dates (one of the days had 2 experiments, thus 17 experiments in total). We processed and extracted the dimensional properties and velocity trajectory of 2,039 particles from this database. The remaining 1,225 trajectories could not be tracked due to the technical limitations of the image processing and tracking algorithms. For example, due to the sharp refractive index gradient in zone (b) in Figs.~\ref{fig:optical_distortion}-\ref{fig:typical_trajs}, some trajectories were misidentified as two separate trajectories; one starting from above the interface and ending within it, and the other starting somewhere within the interface and continuing until the end of the frame. We did not attempt to match fractions of trajectories together and deleted them from the dataset. Other examples of deleted trajectories are of particles that resulted in a calculated expected time ($\hat{t}_{V_1-V2}$) that is greater than the measured retention time ($t_{V_1-V2}$). By definition $\hat{t}_{V_1-V2} \leq t_{V_1-V2}$ leading to $0 \leq \tau$. However, since we do not measure $\hat{t}_{V_1-V2}$ directly but estimate it using the averaged interfacial layer thickness $\hat{h}$, some particles appear as they have $\hat{t}_{V_1-V2} > t_{V_1-V2}$.  In such cases, the measured retention time is negative, and we discard these from the dataset used for the following stage. In summary, we have a dataset of 2,039 trajectories for which $0 \leq  \tau < \infty $. 

The dimensional properties of all the particles in this dataset are presented in Table.~\ref{tbl:particle_properties}. From this dataset, 18\% of the analyzed particles do not experience a minimum velocity, and 50\% exhibit a minimum that is significantly lower than $V_2$ (more than a 5\% decrease from the terminal velocity in the bottom layer). This leaves 32\% of particles with a minimum that is smaller than $V_2$ by $0.01\%-4.5\%$. Due to the large amount of uncertainty associated with the measurements, it is difficult to determine whether these particles truly experience a minimum or not.  

\begin{table}[!ht]
\centering
\caption{Ranges of the properties of the particles and the dimensionless numbers of the top and bottom layers varied across all experiments.}
\label{tbl:particle_properties}
\begin{tabular}{ p{1.8cm}p{2.6cm}p{2.6cm}p{2cm}p{1.7cm}p{2cm}p{1.3cm} } 
\hline
$a$ & $\rho_p$ & $V_1$ & $V_2$ & $\tau$ & \review{$Re_1$} & \review{$Fr_1$}\\
(mm) & (g cm$^{-3}$) & (cm s$^{-1}$) & (cm s$^{-1}$) & (s) & \review{(-)} & \review{(-)}\\
\hline
1 - 1.18  & 2,425 - 2,575  & 11.06 - 21.48 & 9 - 17.43 & 0 - 0.49 & 116 - 242 & 19 - 54 \\
\end{tabular}
\end{table}
%

The dataset comprises of 9 dimensional properties of particles and fluids, \review{i.e., particle diameter and density $\rho_p,a$, the terminal velocities $V_1,V_2$, the delay time $\tau$, the fluids density and viscosity $\rho_1,\rho_2,\nu$, and the averaged interfacial layer thickness $\hat{h}$}. Prior to symbolic regression of a correlation for $\tau$, we initially create a dataset of non-dimensional parameters. 

According to the Buckingham theorem of dimensional analysis, the number of non-dimensional parameters $\pi_{i}$ equals the number of relevant dimensional variables minus the number of dimensional units. In our case, the number of dimensional units is 3: either mass, length, and time, or force, length, and time. We estimated that we need six non-dimensional parameters to formulate a correlation for the delay time, listed in Table.~\ref{tbl:feature_groups}.

Second, we had to select the form of every dimensionless parameter. This is because every dimensionless parameter $\pi_i$ can be calculated in various options based on variables with the same dimensions (between 1 and 384 different options depending on the complexity of the dimensionless parameter). For instance, the Reynolds-like parameters can be defined using one of the two characteristic length scales ($a, \hat{h}$) and three different velocities ($V_1, V_2, \overline{V} = 0.5(V_1+V_2)$), resulting in six plausible options. Each of these options will lead to a different meaning of the parameter and different relative contributions of the physical properties in that parameter. Table.~\ref{tbl:feature_groups} includes the number of options for each parameter. 

Third, we needed a non-dimensional time scale based on the delay time $\tau$ defined above. For that, we used the expected crossing time $\hat{t}_{V_1-V_2}$ for the normalization. This value can be obtained for the particle of a known size and density for a given interfacial layer. Therefore, it is useful for future prediction applications. The normalized delay time scale is then $\hat{\tau}=\tau / \hat{t}_{V_1-V_2}$. Theoretically, it can be any value in the range $0 \leq \hat{\tau} < \infty$, with infinite values corresponding to particles entrapped in the interfacial layer.

\subsection*{Selection of six non-dimensional parameters through a feature selection process}

The first stage in symbolic regression (SR) was to choose the ``best'' option for each of the six $\pi_i$ parameters (e.g., Reynolds-like, Froude-like, etc.) out of a total of 412 different options presented (see No. of options column in Table~\ref{tbl:feature_groups}). This is a major component of the proposed methodology, called \textit{feature selection}, and it is only available due to the large dataset of measurements. We implemented the process in \scimed, which selects six features (each feature being one of the options for each $\pi_i$) that are ``most informative'' to the dimensionless delay time. ``Most informative'' means that an ML model for ``black box'' prediction, generated the most accurate predictions of $\hat{\tau}$ using the data of the selected six features, compared to all other feature combinations.

During the \textit{feature selection} process, \scimed incorporates the information about the category of each feature, meaning to which $\pi_i$ parameter it can be assigned. Then, it uses a genetic algorithm (GA) to generate tens of thousands of subsets of the original dataset, where each subset includes only six features, one from each $\pi_i$. The subsets are evaluated and compared by training an AutoML model~\cite{scimed} that predicts $\hat{\tau}$ based on the selected set of features. The score of each subset is determined by the accuracy of the prediction that was achieved. Finally, the subset leading to the most accurate result is passed as the dataset for the SR process.

\subsection*{Symbolic regression for the delay time}

In this step, \scimed replaces the scientist, and instead of manually searching for the best fit, it suggests the optimal equation structure and parameters that best predict $\hat{\tau}$ of all particles in the dataset. For that purpose, the \textit{Las Vegas} - based SR component \cite{scimed} searches for the optimal correlation. This step results in a list of three possible equations, ordered by their complexity (in terms of the number of operators and parameters in it). 

Each equation consists of a numerical prefactor $\alpha$ and a constant term $\beta$ that is optimized for our dataset of samples (i.e., $\hat{\tau}=\alpha f(\pi_1,...,\pi_6)+\beta$). For cases of data collected from particles or fluid medium that differ from this experiment, such as the data by \citet{verso2019}, we run another optimizer to obtain the $\alpha$ and $\beta$ numerical terms. Eventually, for the sake of generalization, we suggest one ML-selected correlation (i.e., equation resulting from the SR) that best fits both our data and the data from \citet{verso2019}.

\section*{Results}\label{sec:results}
\subsection*{Feature selection}

We report here the set of six $\pi_i$ that result from a  \textit{feature selection} process, comparing 43,000 variations. This set supposedly comprises the dimensionless numbers representing the dominant mechanisms determining the normalized delay time $\hat{\tau}$. 

In Table~\ref{tbl:feature_groups}, we present a comparison between the most typical choice in the literature and the ML-selected form of each $\pi_i$ parameter. In the right column, we present the value range for each parameter according to the selected form and our measurements. 

\begin{table}[!ht]
\centering
\caption{Comparison of the six non-dimensional parameters in the conventional form and the form selected by \scimed. For the length and velocity scales, there is only one reasonable selection. The ranges are for the selected form.}
\label{tbl:feature_groups}
\begin{tabular}{ p{4cm}p{2.9cm}p{3.6cm}p{2.7cm}l } 
\hline
$\pi_i$  & No. of options & Conventional form & Selected form & Range \\
\hline
Length scales & 1 & $\hat{h}/a$ & $\hat{h}/a$ & 25 - 50 \\
Characteristic velocity & 1 & $V_1/V_2$ & $V_1/V_2$ & 1 - 1.8\\
$Re$-like & 6 & $V_1 a/\nu$ & $\overline{V} a/\nu$ & 103 - 213\\
Characteristic density & 4 & $\rho_2/\rho_1$ & $\rho_p/\overline{\rho}$ & 2.20 - 2.53 \\
Density jump & 16 & $\rho_2/(\rho_2-\rho_1)$ & $\rho_2/(\rho_p-\overline{\rho})$ & 0.67 - 0.91\\
$Fr$-like & 384 & $V_1/a\sqrt{\frac{g}{\hat{h}} \frac{\rho_2-\rho_1}{\overline{\rho}}}$ &  $V_1/h\sqrt{\frac{g}{a} \frac{\rho_p-\rho_2}{\overline{\rho}}}$ & 0.02 - 0.06\\
\end{tabular}
\end{table}

\begin{figure}[!ht]
    \centering
    \includegraphics[width=1.\textwidth]{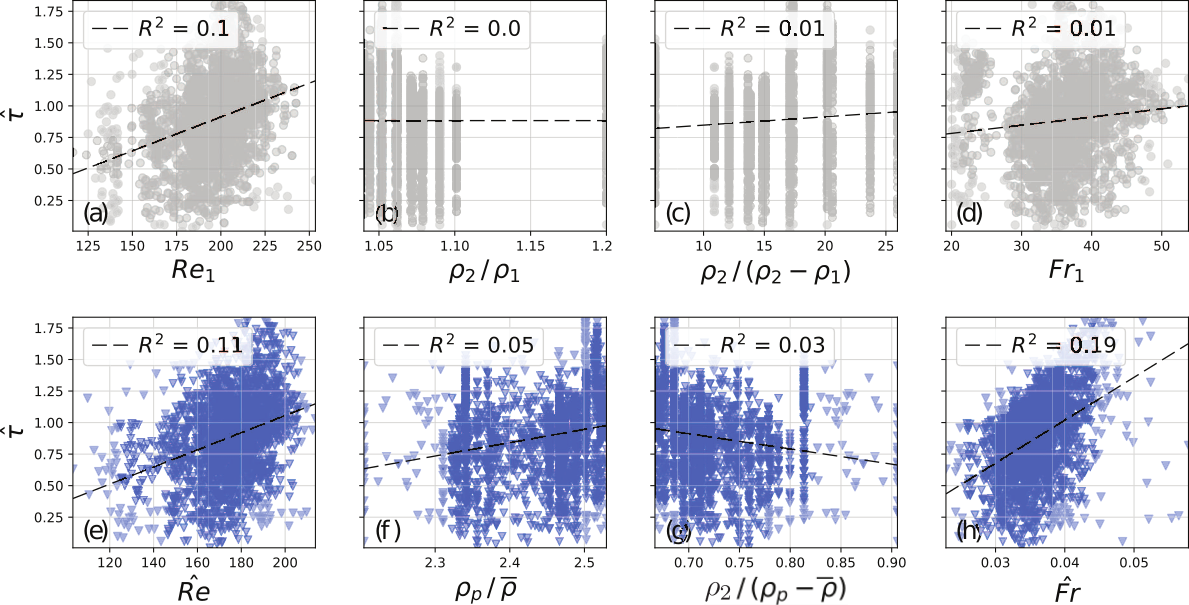}
    \caption{Normalized delay time $\hat{\tau}$ versus four different $\pi_i$ parameters in their conventional form (top row) and the from selected by \scimed (bottom row). \review{Results are shown with the coefficient of determination, $R^2$, representing the variance captured by a linear regression between the X and Y axis of each plot (dashed line)}}
    \label{fig:conv_vs_select}
\end{figure}

In Fig.~\ref{fig:conv_vs_select}, we compare $\hat{\tau}=f(\pi_i)$ in the conventional (top row) and the selected (bottom row) form of four different $\pi_i$ parameters. The dimensionless parameters of length and velocity are not presented in this comparison as, in this case, there is no other plausible form to calculate them. As seen in the figure, the collapse of data of the $\pi_i$ parameters versus the normalized delay time leads to improved correlation, although weak for each dimensionless parameter separately. Furthermore, the parameters leading to the most explicit trends are the selected $Re$-like and $Fr$-like ratios, which from here on now, will be named $\hat{Re}$ and $\hat{Fr}$.

\begin{figure}[!ht]
    \centering
    \includegraphics[width=0.7\textwidth]{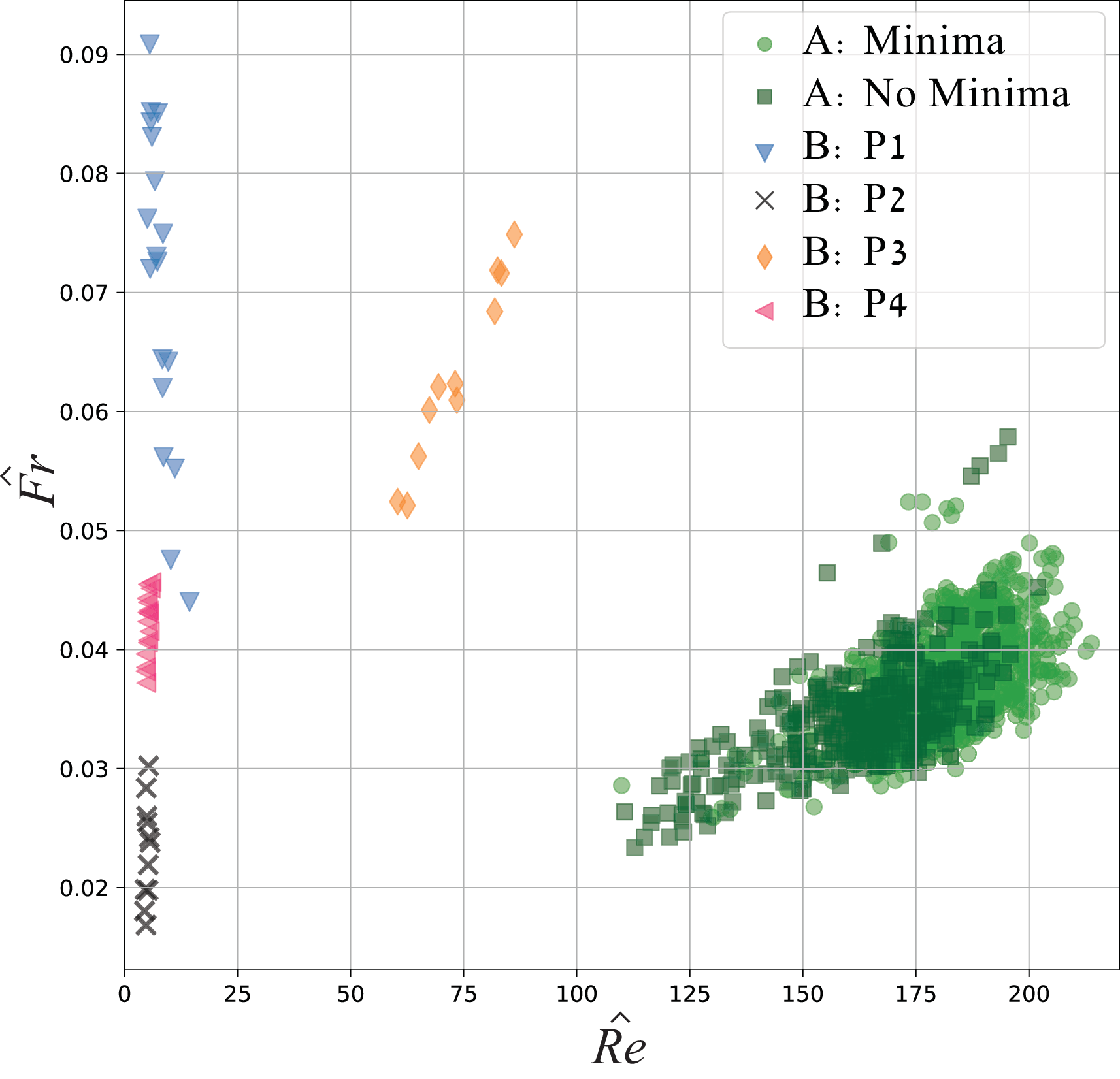}
    \caption{Parameter range of the experimental results of this study (A) and of four types of particles presented in \citet{verso2019} (B P1-P4), in terms of $\hat{Fr}$ versus $\hat{Re}$ (i.e., the $Re$ and $Fr$-like parameters selected by \scimed).}
    \label{fig:new_fr_re}
\end{figure}

In Fig.~\ref{fig:new_fr_re}, we present our measurements (marked as A, for either particle that experiences a significant velocity minimum or those that do not) and that of \citet{verso2019} (marked as B, varying from type P1 to P4), over the $\hat{Fr}=f(\hat{Re})$ map, similarly to Fig.~\ref{fig:Re_Fr_map}. We did not plot the measurements of \citet{Abaid2004} and \citet{fernando1999} that were presented in Fig.~\ref{fig:Re_Fr_map}, as we do not have access to the data needed to calculate $\hat{Fr}$. Note that particle type P3 corresponds to the case that did not show minima (see \citet{verso2019}). This new map emphasizes for the first time that in terms of the new dimensionless parameters $\hat{Fr}$ and $\hat{Re}$, all the particle types are different, a fact that was not observable in the conventional $Fr_1=f(Re_1)$ map~\cite{verso2019}. Note that for our measurements, there is also a notable though the small difference between the particles without and with significant minima. 

\clearpage

\subsection*{Symbolic regression and new correlations}

Table.~\ref{tbl:3_suggested_corr} presents the three types of equations suggested by the SR component of \scimed together with the numerical constants optimized to our dataset. The three typically used success metrics of \review{coefficient of determination} $R^2$, mean absolute error (MAE), and mean squared error (MSE) presented were calculated from all samples of our study, together with those of \citet{verso2019}. As seen in the table, all the suggested correlations resulted in similar scores. It is difficult to decide which correlation is ``better,'' but this is also not the purpose of this study. In order to obtain a much better correlation, we need to obtain more measurements of very different types of particles, with different fluids, interfacial layer thickness, and density jumps. To this end, we obtained symbolic expressions representing the result better than the previous studies. Furthermore, we focus on the question of why these correlations are better than the previous ones and what underlying physical mechanisms these selections could highlight. 

\begin{table}[!ht]
    \centering
    \caption{Three plausible correlations for the normalized delay time, as suggested by \scimed. The $R^2$, mean absolute error (MAE), and mean squared error (MSE) metric scores are reported based on measurements of this study and that of \citet{verso2019}.}
    \begin{tabular}{p{1cm}p{5.6cm}p{4.4cm}p{1.3cm}p{1.3cm}p{1cm}}
    \hline
           &  Equation  & Constants & $R^2$ & MAE & MSE  \\
    \hline
    $f_1$  &  $\alpha_1 \big( \hat{Fr} \frac{\rho_p}{\overline{\rho}} \big)+\beta_1$ & $\alpha_1=16.19,\; \beta_1=-0.5$ & $0.75$ & 0.24 & 0.23\\
    $f_2$  &  $\alpha_2 \big( \hat{Fr} \hat{Re}  \frac{h}{a} \big)+\beta_2$ & $\alpha_2=2.76,\; \beta_2=0.47$ & $0.71$ & 0.23 & 0.20 \\
    $f_3$  &  $\alpha_3 \big( \hat{Fr}^2  \frac{\rho_p}{\overline{\rho}} (1+\frac{\rho_p}{\overline{\rho}} \frac{V_2}{V_1}) \big)+\beta_3$ & $\alpha_3=89.91,\; \beta_3=0.44$ & $0.72$ & 0.24 & 0.23  
    \end{tabular}
    \label{tbl:3_suggested_corr}
\end{table} 

\begin{figure}
    \centering
    \includegraphics[width=1.\textwidth]{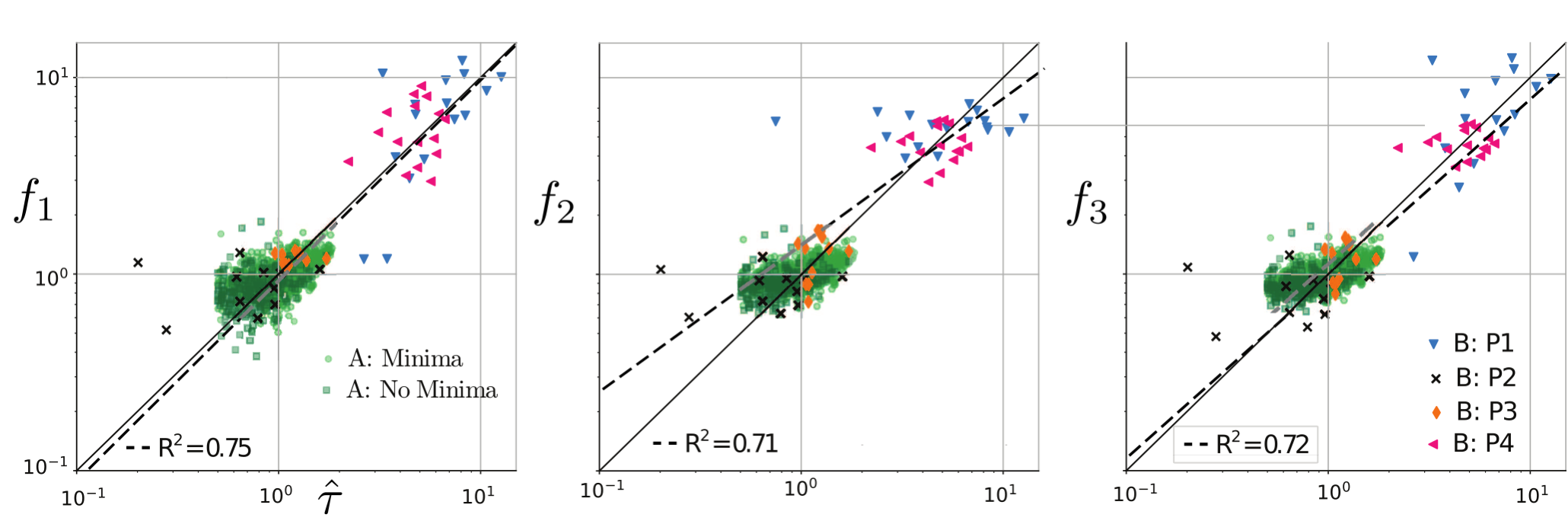}
    \caption{Predicted normalized delay time as calculated by $f_1-f_3$ versus the actual measurements of $\hat{\tau}$. The scatter is the experimental results of this study (A) and of four types of particles presented in \citet{verso2019} (B P1-P4).}
    \label{fig:3_suggested_corr}
\end{figure}

In Fig.~\ref{fig:3_suggested_corr}, we show the scatter of the predicted normalized delay time (using $f_1-f_3$) versus the actual measurements of $\hat{\tau}$. In each plot, a linear trend line (dashed line) is fitted to all predicted samples. It is clear that the trend line fitted to the predictions of $f_1$ (left) is closest to the expected $y=x$ trend (solid line). It is also noteworthy that the results were obtained with particles measured in two separate sets of experiments (ours and that of \cite{verso2019}), using two different fluid layer combinations, five particle types, and 10 density jumps. Based on them we suggest to use $f_1$ as the correlation for $\hat{\tau}$, which can be written as:

\begin{eqnarray}\label{eq:correlation}
    \hat{N} &=& \sqrt{\frac{g}{a} \frac{\rho_p-\rho_2}{\overline{\rho}}} \\
    f_1 &=& \alpha \frac{\rho_p}{\overline{\rho}} \, \frac{V_1}{\hat{h} \hat{N}} + \beta 
\end{eqnarray}
and if we substitute $\hat{N}$ it reads:
\begin{equation}
    \label{eq:correlation2}
    f_1 = \alpha \frac{V_1}{h} \sqrt{\frac{a}{g} \frac{\rho_p}{\rho_p-\rho_2} \frac{\rho_p}{\overline{\rho}}} + \beta 
\end{equation}

\section*{Discussion}\label{sec:discussion}

In this study, we took a new approach to reveal the structure of empirical results. We utilized machine learning techniques to study the complex behavior of inertial particles crossing an interfacial stratified layer. We simplified the experimental setup to gain a substantial expansion of the previously unexplored parameter range and an unprecedented number of particle trajectory datasets. With this data, we aimed to find a correlation that covers our and previous results. All the experimental results are open to the community, and we hope that additional ideas and new correlations can be found through new approaches. 

Consistently with previous results~\cite{fernando1999, Abaid2004, mandel2020, verso2019}, we also find that not all particles in this parameter range experience slowdown and minimal velocity. In our case, 18\% of all the trajectories did not experience a minima. More research is needed to understand why this effect occurs, but possible factors include particle rotation, the timing of particle release, non-spherical shapes, or interface oscillation. New experiments would be necessary to better understand this effect.

We demonstrate that the normalized delay time ($\hat{\tau}$) has a stronger dependence on the newly selected forms of dimensionless parameters rather than the conventional form, as shown in Fig.~\ref{fig:conv_vs_select}. We also inspect the selected forms of the dimensionless parameters and infer the possible meaning of the selection. Thus, $\hat{Re}$ is based on average velocity and not on the entrance velocity as it was suggested by previous works~\cite{fernando1999, verso2019}. The Reynolds number does depend on the particle size, and it represents the so-called particle Reynolds number. 

The particle density is normalized to the averaged fluid density (which is the approximate density of the interface, assuming linear density gradient). It seems to be more reasonable to incorporate both densities instead of the conventional $\rho_2/\rho_1$.

We also learn that because there is a strong correlation between the Reynolds and Froude numbers (see the new map in Fig.~\ref{fig:new_fr_re}), the final correlation does not include the Reynolds number explicitly. The most dominant effect in this problem is stratification, and it determines the retention and the delay times. Therefore, the most dominant dimensionless parameter here is the Froude-like parameter. Furthermore, the newly selected Froude number differs from the one typically suggested in the literature. It relates linearly to the interfacial layer thickness, $h$, and only as a square root of the particle diameter $a^{-0.5}$. In other words, we can infer that the dimensionless frequency, $\hat{N}$ should be defined with the particle diameter, while Froude number as $\hat{Fr} = V/\hat{h}\,\hat{N}$. The better choice of this form for the Froude number is also supported by the better collapse of data than the rest of the $\pi_i$ parameters. We also compared it to two Froude number forms and see that it is significantly stronger related to the delay time, i.e. ($\hat{\tau}=f(\hat{Fr})$ leads to $R^2=0.19$, while both $Fr_1$ and $Fr_2$ lead to an order of magnitude weaker correlation with $R^2=0.01$.

Undoubtedly, our experiment has limitations. Although it enabled the sampling of thousands of trajectories, it also led to increased uncertainty and lesser control over particle diameter, density, and less precise release timing. The main disadvantage is the lack of index refraction matching and the corresponding lack of detailed trajectories inside the interfacial layer. If such an experiment can be performed, it would be possible to implement \scimed for particle trajectories and compare the point-wise position and velocity of each particle versus the predicted equation of motion. We expect this approach to significantly improve our ability to predict the correct form of the stratification force and formulate a model in a more general form of detailed physical mechanisms. \review{Additionally, we appreciate the idea of an anonymous reviewer to try to construct additional dimensionless features using a new length scale that instead of} \(h\) will be of the type of $l_\rho = 2 ( \partial \rho / \partial z )( \partial^2 \rho / \partial z^2)$. \review{We encourage the readers to use SciMED’s open-source code and the openly shared data to verify whether this analysis leads to a better prediction of the retention time scale.}

\clearpage

\section*{Acknowledgments}
The authors thank Daniel Jalontzki for collecting the settling particle data, Aviv Littman for measuring density profiles, and Dmitry Aronovsky for the experimental setup and assistance with experiments. 

\section*{Declarations}
\subsection*{Funding}
This research was partially supported by Israel Science Foundation (grant number 441/2) and Gordon center for energy research at Tel Aviv University.

\subsection*{Data availability}
The data in this study is available in the manuscript with the relevant sources.

\subsection*{Conflicts of interest}
The authors have no financial or proprietary interests in any material discussed in this article.

\bibliography{library}

\begin{thebibliography}{25}%
\makeatletter
\providecommand \@ifxundefined [1]{%
 \@ifx{#1\undefined}
}%
\providecommand \@ifnum [1]{%
 \ifnum #1\expandafter \@firstoftwo
 \else \expandafter \@secondoftwo
 \fi
}%
\providecommand \@ifx [1]{%
 \ifx #1\expandafter \@firstoftwo
 \else \expandafter \@secondoftwo
 \fi
}%
\providecommand \natexlab [1]{#1}%
\providecommand \enquote  [1]{``#1''}%
\providecommand \bibnamefont  [1]{#1}%
\providecommand \bibfnamefont [1]{#1}%
\providecommand \citenamefont [1]{#1}%
\providecommand \href@noop [0]{\@secondoftwo}%
\providecommand \href [0]{\begingroup \@sanitize@url \@href}%
\providecommand \@href[1]{\@@startlink{#1}\@@href}%
\providecommand \@@href[1]{\endgroup#1\@@endlink}%
\providecommand \@sanitize@url [0]{\catcode `\\12\catcode `\$12\catcode
  `\&12\catcode `\#12\catcode `\^12\catcode `\_12\catcode `\%12\relax}%
\providecommand \@@startlink[1]{}%
\providecommand \@@endlink[0]{}%
\providecommand \url  [0]{\begingroup\@sanitize@url \@url }%
\providecommand \@url [1]{\endgroup\@href {#1}{\urlprefix }}%
\providecommand \urlprefix  [0]{URL }%
\providecommand \Eprint [0]{\href }%
\providecommand \doibase [0]{https://doi.org/}%
\providecommand \selectlanguage [0]{\@gobble}%
\providecommand \bibinfo  [0]{\@secondoftwo}%
\providecommand \bibfield  [0]{\@secondoftwo}%
\providecommand \translation [1]{[#1]}%
\providecommand \BibitemOpen [0]{}%
\providecommand \bibitemStop [0]{}%
\providecommand \bibitemNoStop [0]{.\EOS\space}%
\providecommand \EOS [0]{\spacefactor3000\relax}%
\providecommand \BibitemShut  [1]{\csname bibitem#1\endcsname}%
\let\auto@bib@innerbib\@empty
\bibitem [{\citenamefont {Verso}\ \emph {et~al.}(2019)\citenamefont {Verso},
  \citenamefont {van Reeuwijk},\ and\ \citenamefont {Liberzon}}]{verso2019}%
  \BibitemOpen
  \bibfield  {author} {\bibinfo {author} {\bibfnamefont {L.}~\bibnamefont
  {Verso}}, \bibinfo {author} {\bibfnamefont {M.}~\bibnamefont {van
  Reeuwijk}},\ and\ \bibinfo {author} {\bibfnamefont {A.}~\bibnamefont
  {Liberzon}},\ }\bibfield  {title} {\bibinfo {title} {Transient stratification
  force on particles crossing a density interface},\ }\href@noop {} {\bibfield
  {journal} {\bibinfo  {journal} {International Journal of Multiphase Flow}\
  }\textbf {\bibinfo {volume} {121}},\ \bibinfo {pages} {103109} (\bibinfo
  {year} {2019})}\BibitemShut {NoStop}%
\bibitem [{\citenamefont {Magnaudet}\ and\ \citenamefont
  {Mercier}(2020)}]{magnaudet2020}%
  \BibitemOpen
  \bibfield  {author} {\bibinfo {author} {\bibfnamefont {J.}~\bibnamefont
  {Magnaudet}}\ and\ \bibinfo {author} {\bibfnamefont {M.}~\bibnamefont
  {Mercier}},\ }\bibfield  {title} {\bibinfo {title} {Particles, drops, and
  bubbles moving across sharp interfaces and stratified layers},\ }\href@noop
  {} {\bibfield  {journal} {\bibinfo  {journal} {Annual Review of Fluid
  Mechanics}\ }\textbf {\bibinfo {volume} {52}},\ \bibinfo {pages} {61}
  (\bibinfo {year} {2020})}\BibitemShut {NoStop}%
\bibitem [{\citenamefont {Mrokowska}(2018)}]{Mrokowska2018}%
  \BibitemOpen
  \bibfield  {author} {\bibinfo {author} {\bibfnamefont {M.~M.}\ \bibnamefont
  {Mrokowska}},\ }\bibfield  {title} {\bibinfo {title} {{Stratification-induced
  reorientation of disk settling through ambient density transition}},\ }\href
  {https://doi.org/10.1038/s41598-017-18654-7} {\bibfield  {journal} {\bibinfo
  {journal} {Scientific Reports}\ }\textbf {\bibinfo {volume} {8}},\ \bibinfo
  {pages} {1} (\bibinfo {year} {2018})}\BibitemShut {NoStop}%
\bibitem [{\citenamefont {Renggli}\ \emph {et~al.}(2016)\citenamefont
  {Renggli}, \citenamefont {Wiesmaier}, \citenamefont {De~Campos},
  \citenamefont {Hess},\ and\ \citenamefont
  {Dingwell}}]{Renggli2016MagmaSettling}%
  \BibitemOpen
  \bibfield  {author} {\bibinfo {author} {\bibfnamefont {C.~J.}\ \bibnamefont
  {Renggli}}, \bibinfo {author} {\bibfnamefont {S.}~\bibnamefont {Wiesmaier}},
  \bibinfo {author} {\bibfnamefont {C.~P.}\ \bibnamefont {De~Campos}}, \bibinfo
  {author} {\bibfnamefont {K.~U.}\ \bibnamefont {Hess}},\ and\ \bibinfo
  {author} {\bibfnamefont {D.~B.}\ \bibnamefont {Dingwell}},\ }\bibfield
  {title} {\bibinfo {title} {{Magma mixing induced by particle settling}},\
  }\href@noop {} {\bibfield  {journal} {\bibinfo  {journal} {Contributions to
  Mineralogy and Petrology}\ }\textbf {\bibinfo {volume} {171}},\ \bibinfo
  {pages} {96} (\bibinfo {year} {2016})}\BibitemShut {NoStop}%
\bibitem [{\citenamefont {Verso}(2020)}]{verso2020}%
  \BibitemOpen
  \bibfield  {author} {\bibinfo {author} {\bibfnamefont {L.}~\bibnamefont
  {Verso}},\ }\emph {\bibinfo {title} {Particles crossing density
  interfaces}},\ \href@noop {} {Ph.D. thesis},\ \bibinfo  {school} {Tel Aviv
  University} (\bibinfo {year} {2020})\BibitemShut {NoStop}%
\bibitem [{\citenamefont {Abaid}\ \emph {et~al.}(2004)\citenamefont {Abaid},
  \citenamefont {Adalsteinsson}, \citenamefont {Agyapong},\ and\ \citenamefont
  {McLaughlin}}]{Abaid2004}%
  \BibitemOpen
  \bibfield  {author} {\bibinfo {author} {\bibfnamefont {N.}~\bibnamefont
  {Abaid}}, \bibinfo {author} {\bibfnamefont {D.}~\bibnamefont
  {Adalsteinsson}}, \bibinfo {author} {\bibfnamefont {A.}~\bibnamefont
  {Agyapong}},\ and\ \bibinfo {author} {\bibfnamefont {R.}~\bibnamefont
  {McLaughlin}},\ }\bibfield  {title} {\bibinfo {title} {An internal splash:
  Levitation of falling spheres in stratified fluids},\ }\href@noop {}
  {\bibfield  {journal} {\bibinfo  {journal} {Physics of Fluids}\ }\textbf
  {\bibinfo {volume} {16}},\ \bibinfo {pages} {1567} (\bibinfo {year}
  {2004})}\BibitemShut {NoStop}%
\bibitem [{\citenamefont {Srdi\'{c}-Mitrovi\'{c}}\ \emph
  {et~al.}(1999)\citenamefont {Srdi\'{c}-Mitrovi\'{c}}, \citenamefont
  {Mohamed},\ and\ \citenamefont {Fernando}}]{fernando1999}%
  \BibitemOpen
  \bibfield  {author} {\bibinfo {author} {\bibfnamefont {A.}~\bibnamefont
  {Srdi\'{c}-Mitrovi\'{c}}}, \bibinfo {author} {\bibfnamefont {N.}~\bibnamefont
  {Mohamed}},\ and\ \bibinfo {author} {\bibfnamefont {H.}~\bibnamefont
  {Fernando}},\ }\bibfield  {title} {\bibinfo {title} {Gravitational settling
  of particles through density interfaces},\ }\href
  {https://doi.org/10.1017/S0022112098003590} {\bibfield  {journal} {\bibinfo
  {journal} {Journal of Fluid Mechanics}\ }\textbf {\bibinfo {volume} {381}},\
  \bibinfo {pages} {175–198} (\bibinfo {year} {1999})}\BibitemShut {NoStop}%
\bibitem [{\citenamefont {Mandel}\ \emph {et~al.}(2020)\citenamefont {Mandel},
  \citenamefont {Waldrop}, \citenamefont {Theillard}, \citenamefont {Kleckner},
  \citenamefont {Khatri} \emph {et~al.}}]{mandel2020}%
  \BibitemOpen
  \bibfield  {author} {\bibinfo {author} {\bibfnamefont {T.}~\bibnamefont
  {Mandel}}, \bibinfo {author} {\bibfnamefont {L.}~\bibnamefont {Waldrop}},
  \bibinfo {author} {\bibfnamefont {M.}~\bibnamefont {Theillard}}, \bibinfo
  {author} {\bibfnamefont {D.}~\bibnamefont {Kleckner}}, \bibinfo {author}
  {\bibfnamefont {S.}~\bibnamefont {Khatri}}, \emph {et~al.},\ }\bibfield
  {title} {\bibinfo {title} {Retention of rising droplets in density
  stratification},\ }\href@noop {} {\bibfield  {journal} {\bibinfo  {journal}
  {Physical Review Fluids}\ }\textbf {\bibinfo {volume} {5}},\ \bibinfo {pages}
  {124803} (\bibinfo {year} {2020})}\BibitemShut {NoStop}%
\bibitem [{\citenamefont {MacIntyre}\ \emph {et~al.}(1995)\citenamefont
  {MacIntyre}, \citenamefont {Alldredge},\ and\ \citenamefont
  {Gotschalk}}]{MacIntyre1995}%
  \BibitemOpen
  \bibfield  {author} {\bibinfo {author} {\bibfnamefont {S.}~\bibnamefont
  {MacIntyre}}, \bibinfo {author} {\bibfnamefont {A.}~\bibnamefont
  {Alldredge}},\ and\ \bibinfo {author} {\bibfnamefont {C.}~\bibnamefont
  {Gotschalk}},\ }\bibfield  {title} {\bibinfo {title} {Accumulation of marine
  snow at density discontinuities in the water column},\ }\href@noop {}
  {\bibfield  {journal} {\bibinfo  {journal} {Limnology and Oceanography}\
  }\textbf {\bibinfo {volume} {40}},\ \bibinfo {pages} {449} (\bibinfo {year}
  {1995})}\BibitemShut {NoStop}%
\bibitem [{\citenamefont {Prairie}\ \emph {et~al.}(2013)\citenamefont
  {Prairie}, \citenamefont {Ziervogel}, \citenamefont {Arnosti}, \citenamefont
  {Camassa}, \citenamefont {Falcon}, \citenamefont {Khatri}, \citenamefont
  {McLaughlin}, \citenamefont {White},\ and\ \citenamefont {Yu}}]{Prairie2013}%
  \BibitemOpen
  \bibfield  {author} {\bibinfo {author} {\bibfnamefont {J.~C.}\ \bibnamefont
  {Prairie}}, \bibinfo {author} {\bibfnamefont {K.}~\bibnamefont {Ziervogel}},
  \bibinfo {author} {\bibfnamefont {C.}~\bibnamefont {Arnosti}}, \bibinfo
  {author} {\bibfnamefont {R.}~\bibnamefont {Camassa}}, \bibinfo {author}
  {\bibfnamefont {C.}~\bibnamefont {Falcon}}, \bibinfo {author} {\bibfnamefont
  {S.}~\bibnamefont {Khatri}}, \bibinfo {author} {\bibfnamefont {R.~M.}\
  \bibnamefont {McLaughlin}}, \bibinfo {author} {\bibfnamefont {B.~L.}\
  \bibnamefont {White}},\ and\ \bibinfo {author} {\bibfnamefont
  {S.}~\bibnamefont {Yu}},\ }\bibfield  {title} {\bibinfo {title} {{Delayed
  settling of marine snow at sharp density transitions driven by fluid
  entrainment and diffusion-limited retention}},\ }\href
  {https://doi.org/10.3354/meps10387} {\bibfield  {journal} {\bibinfo
  {journal} {Marine Ecology Progress Series}\ }\textbf {\bibinfo {volume}
  {487}},\ \bibinfo {pages} {185} (\bibinfo {year} {2013})}\BibitemShut
  {NoStop}%
\bibitem [{\citenamefont {Turco}\ \emph {et~al.}(1983)\citenamefont {Turco},
  \citenamefont {Toon}, \citenamefont {Ackerman}, \citenamefont {Pollack},\
  and\ \citenamefont {Sagan}}]{Turco1983}%
  \BibitemOpen
  \bibfield  {author} {\bibinfo {author} {\bibfnamefont {R.}~\bibnamefont
  {Turco}}, \bibinfo {author} {\bibfnamefont {O.}~\bibnamefont {Toon}},
  \bibinfo {author} {\bibfnamefont {T.}~\bibnamefont {Ackerman}}, \bibinfo
  {author} {\bibfnamefont {J.}~\bibnamefont {Pollack}},\ and\ \bibinfo {author}
  {\bibfnamefont {C.}~\bibnamefont {Sagan}},\ }\bibfield  {title} {\bibinfo
  {title} {Nuclear winter: Global consequences of multiple nuclear
  explosions},\ }\href@noop {} {\bibfield  {journal} {\bibinfo  {journal}
  {Science}\ }\textbf {\bibinfo {volume} {222}},\ \bibinfo {pages} {1283}
  (\bibinfo {year} {1983})}\BibitemShut {NoStop}%
\bibitem [{\citenamefont {Kok}(2011)}]{kok2011}%
  \BibitemOpen
  \bibfield  {author} {\bibinfo {author} {\bibfnamefont {J.}~\bibnamefont
  {Kok}},\ }\bibfield  {title} {\bibinfo {title} {A scaling theory for the size
  distribution of emitted dust aerosols suggests climate models underestimate
  the size of the global dust cycle},\ }\href@noop {} {\bibfield  {journal}
  {\bibinfo  {journal} {Proceedings of the National Academy of Sciences}\
  }\textbf {\bibinfo {volume} {108}},\ \bibinfo {pages} {1016} (\bibinfo {year}
  {2011})}\BibitemShut {NoStop}%
\bibitem [{\citenamefont {Smith}\ \emph {et~al.}(1992)\citenamefont {Smith},
  \citenamefont {Simon}, \citenamefont {Alldredge},\ and\ \citenamefont
  {Azam}}]{smith1992}%
  \BibitemOpen
  \bibfield  {author} {\bibinfo {author} {\bibfnamefont {D.}~\bibnamefont
  {Smith}}, \bibinfo {author} {\bibfnamefont {M.}~\bibnamefont {Simon}},
  \bibinfo {author} {\bibfnamefont {A.}~\bibnamefont {Alldredge}},\ and\
  \bibinfo {author} {\bibfnamefont {F.}~\bibnamefont {Azam}},\ }\bibfield
  {title} {\bibinfo {title} {Intense hydrolytic enzyme activity on marine
  aggregates and implications for rapid particle dissolution},\ }\href@noop {}
  {\bibfield  {journal} {\bibinfo  {journal} {Nature}\ }\textbf {\bibinfo
  {volume} {359}},\ \bibinfo {pages} {139} (\bibinfo {year}
  {1992})}\BibitemShut {NoStop}%
\bibitem [{\citenamefont {Burd}\ and\ \citenamefont
  {Jackson}(2009)}]{burd2009}%
  \BibitemOpen
  \bibfield  {author} {\bibinfo {author} {\bibfnamefont {A.}~\bibnamefont
  {Burd}}\ and\ \bibinfo {author} {\bibfnamefont {G.}~\bibnamefont {Jackson}},\
  }\bibfield  {title} {\bibinfo {title} {Particle aggregation},\ }\href@noop {}
  {\bibfield  {journal} {\bibinfo  {journal} {Annual review of marine science}\
  }\textbf {\bibinfo {volume} {1}},\ \bibinfo {pages} {65} (\bibinfo {year}
  {2009})}\BibitemShut {NoStop}%
\bibitem [{\citenamefont {Wang}\ \emph {et~al.}(2023)\citenamefont {Wang},
  \citenamefont {Kandel}, \citenamefont {Deng}, \citenamefont {Caulfield},\
  and\ \citenamefont {Dalziel}}]{Wang2023}%
  \BibitemOpen
  \bibfield  {author} {\bibinfo {author} {\bibfnamefont {S.}~\bibnamefont
  {Wang}}, \bibinfo {author} {\bibfnamefont {P.}~\bibnamefont {Kandel}},
  \bibinfo {author} {\bibfnamefont {J.}~\bibnamefont {Deng}}, \bibinfo {author}
  {\bibfnamefont {C.}~\bibnamefont {Caulfield}},\ and\ \bibinfo {author}
  {\bibfnamefont {S.}~\bibnamefont {Dalziel}},\ }\href
  {https://doi.org/10.48550/arXiv.2301.01484} {} (\bibinfo {year}
  {2023})\BibitemShut {NoStop}%
\bibitem [{\citenamefont {Blanchette}\ and\ \citenamefont
  {Shapiro}(2012)}]{blanchette2012}%
  \BibitemOpen
  \bibfield  {author} {\bibinfo {author} {\bibfnamefont {F.}~\bibnamefont
  {Blanchette}}\ and\ \bibinfo {author} {\bibfnamefont {A.}~\bibnamefont
  {Shapiro}},\ }\bibfield  {title} {\bibinfo {title} {Drops settling in sharp
  stratification with and without marangoni effects},\ }\href@noop {}
  {\bibfield  {journal} {\bibinfo  {journal} {Physics of Fluids}\ }\textbf
  {\bibinfo {volume} {24}},\ \bibinfo {pages} {042104} (\bibinfo {year}
  {2012})}\BibitemShut {NoStop}%
\bibitem [{\citenamefont {Doostmohammadi}\ \emph {et~al.}(2014)\citenamefont
  {Doostmohammadi}, \citenamefont {Dabiri},\ and\ \citenamefont
  {Ardekani}}]{Doostmohammadi2014}%
  \BibitemOpen
  \bibfield  {author} {\bibinfo {author} {\bibfnamefont {A.}~\bibnamefont
  {Doostmohammadi}}, \bibinfo {author} {\bibfnamefont {S.}~\bibnamefont
  {Dabiri}},\ and\ \bibinfo {author} {\bibfnamefont {A.~M.}\ \bibnamefont
  {Ardekani}},\ }\bibfield  {title} {\bibinfo {title} {A numerical study of the
  dynamics of a particle settling at moderate reynolds numbers in a linearly
  stratified fluid},\ }\href@noop {} {\bibfield  {journal} {\bibinfo  {journal}
  {Journal of Fluid Mechanics}\ }\textbf {\bibinfo {volume} {750}},\ \bibinfo
  {pages} {5} (\bibinfo {year} {2014})}\BibitemShut {NoStop}%
\bibitem [{\citenamefont {Yick}\ \emph {et~al.}(2009)\citenamefont {Yick},
  \citenamefont {Torres}, \citenamefont {Peacock},\ and\ \citenamefont
  {Stocker}}]{Yick2009}%
  \BibitemOpen
  \bibfield  {author} {\bibinfo {author} {\bibfnamefont {K.}~\bibnamefont
  {Yick}}, \bibinfo {author} {\bibfnamefont {C.}~\bibnamefont {Torres}},
  \bibinfo {author} {\bibfnamefont {T.}~\bibnamefont {Peacock}},\ and\ \bibinfo
  {author} {\bibfnamefont {R.}~\bibnamefont {Stocker}},\ }\bibfield  {title}
  {\bibinfo {title} {Enhanced drag of a sphere settling in a stratified fluid
  at small reynolds numbers},\ }\href@noop {} {\bibfield  {journal} {\bibinfo
  {journal} {Journal of Fluid Mechanics}\ }\textbf {\bibinfo {volume} {632}},\
  \bibinfo {pages} {49} (\bibinfo {year} {2009})}\BibitemShut {NoStop}%
\bibitem [{\citenamefont {Camassa}\ \emph {et~al.}(2010)\citenamefont
  {Camassa}, \citenamefont {Falcon}, \citenamefont {Lin}, \citenamefont
  {McLaughlin},\ and\ \citenamefont {Mykins}}]{Camassa2010}%
  \BibitemOpen
  \bibfield  {author} {\bibinfo {author} {\bibfnamefont {R.}~\bibnamefont
  {Camassa}}, \bibinfo {author} {\bibfnamefont {C.}~\bibnamefont {Falcon}},
  \bibinfo {author} {\bibfnamefont {J.}~\bibnamefont {Lin}}, \bibinfo {author}
  {\bibfnamefont {R.~M.}\ \bibnamefont {McLaughlin}},\ and\ \bibinfo {author}
  {\bibfnamefont {N.}~\bibnamefont {Mykins}},\ }\bibfield  {title} {\bibinfo
  {title} {{A first-principle predictive theory for a sphere falling through
  sharply stratified fluid at low Reynolds number}},\ }\href
  {https://doi.org/10.1017/S0022112010003800} {\bibfield  {journal} {\bibinfo
  {journal} {Journal of Fluid Mechanics}\ }\textbf {\bibinfo {volume} {664}},\
  \bibinfo {pages} {436} (\bibinfo {year} {2010})}\BibitemShut {NoStop}%
\bibitem [{\citenamefont {Simon~Keren}\ \emph {et~al.}(2023)\citenamefont
  {Simon~Keren}, \citenamefont {Liberzon},\ and\ \citenamefont
  {Lazebnik}}]{scimed}%
  \BibitemOpen
  \bibfield  {author} {\bibinfo {author} {\bibfnamefont {L.}~\bibnamefont
  {Simon~Keren}}, \bibinfo {author} {\bibfnamefont {A.}~\bibnamefont
  {Liberzon}},\ and\ \bibinfo {author} {\bibfnamefont {T.}~\bibnamefont
  {Lazebnik}},\ }\bibfield  {title} {\bibinfo {title} {A computational
  framework for physics-informed symbolic regression with straightforward
  integration of domain knowledge},\ }\href@noop {} {\bibfield  {journal}
  {\bibinfo  {journal} {Scientific Reports}\ }\textbf {\bibinfo {volume}
  {13}},\ \bibinfo {pages} {1249} (\bibinfo {year} {2023})}\BibitemShut
  {NoStop}%
\bibitem [{\citenamefont {Verso}\ \emph {et~al.}(2017)\citenamefont {Verso},
  \citenamefont {van Reeuwijk},\ and\ \citenamefont {Liberzon}}]{verso2017}%
  \BibitemOpen
  \bibfield  {author} {\bibinfo {author} {\bibfnamefont {L.}~\bibnamefont
  {Verso}}, \bibinfo {author} {\bibfnamefont {M.}~\bibnamefont {van
  Reeuwijk}},\ and\ \bibinfo {author} {\bibfnamefont {A.}~\bibnamefont
  {Liberzon}},\ }\bibfield  {title} {\bibinfo {title} {Steady state model and
  experiment for an oscillating grid turbulent two-layer stratified flow},\
  }\href@noop {} {\bibfield  {journal} {\bibinfo  {journal} {Physical Review
  Fluids}\ }\textbf {\bibinfo {volume} {2}},\ \bibinfo {pages} {104605}
  (\bibinfo {year} {2017})}\BibitemShut {NoStop}%
\bibitem [{\citenamefont {{OpenPTV consortium}}(2014)}]{openptv}%
  \BibitemOpen
  \bibfield  {author} {\bibinfo {author} {\bibnamefont {{OpenPTV
  consortium}}},\ }\href {http://www.openptv.net/} {\bibinfo {title} {Open
  source particle tracking velocimetry}} (\bibinfo {year} {2014})\BibitemShut
  {NoStop}%
\bibitem [{\citenamefont {Alahyari}\ and\ \citenamefont
  {Longmire}(1994)}]{alahyari1994particle}%
  \BibitemOpen
  \bibfield  {author} {\bibinfo {author} {\bibfnamefont {A.}~\bibnamefont
  {Alahyari}}\ and\ \bibinfo {author} {\bibfnamefont {E.}~\bibnamefont
  {Longmire}},\ }\bibfield  {title} {\bibinfo {title} {Particle image
  velocimetry in a variable density flow: application to a dynamically evolving
  microburst},\ }\href@noop {} {\bibfield  {journal} {\bibinfo  {journal}
  {Experiments in Fluids}\ }\textbf {\bibinfo {volume} {17}},\ \bibinfo {pages}
  {434} (\bibinfo {year} {1994})}\BibitemShut {NoStop}%
\bibitem [{\citenamefont {White}(1974)}]{white1974viscous}%
  \BibitemOpen
  \bibfield  {author} {\bibinfo {author} {\bibfnamefont {F.}~\bibnamefont
  {White}},\ }\bibfield  {title} {\bibinfo {title} {Viscous fluid flow
  mcgraw-hill inc},\ }\href@noop {} {\bibfield  {journal} {\bibinfo  {journal}
  {New York, New York}\ } (\bibinfo {year} {1974})}\BibitemShut {NoStop}%
\bibitem [{\citenamefont {Boetti}\ and\ \citenamefont
  {Verso}(2022)}]{boetti_verso_2022}%
  \BibitemOpen
  \bibfield  {author} {\bibinfo {author} {\bibfnamefont {M.}~\bibnamefont
  {Boetti}}\ and\ \bibinfo {author} {\bibfnamefont {L.}~\bibnamefont {Verso}},\
  }\bibfield  {title} {\bibinfo {title} {Force on inertial particles crossing a
  two layer stratified turbulent/non-turbulent interface},\ }\href
  {https://doi.org/https://doi.org/10.1016/j.ijmultiphaseflow.2022.104153}
  {\bibfield  {journal} {\bibinfo  {journal} {International Journal of
  Multiphase Flow}\ ,\ \bibinfo {pages} {104153}} (\bibinfo {year}
  {2022})}\BibitemShut {NoStop}%
\end{thebibliography}%
\end{document}